\documentclass[journal]{IEEEtran}
\usepackage{amsmath}
\usepackage{eqparbox}
\usepackage[usenames, dvipsnames]{color}
\usepackage{mathtools,amssymb,bm,mathabx}
\usepackage{amstext}
\usepackage{amssymb}
\usepackage{graphicx}

\usepackage{color}
\usepackage{booktabs}
\usepackage{longtable}
\usepackage{multicol}
\usepackage{amsfonts}
\usepackage{dsfont}
\usepackage{array}
\usepackage[ruled]{algorithm}
\usepackage{algorithmic}
\usepackage{cases}
\usepackage{pgfplots}
\usepackage{caption}
\usepackage{url}
\usepackage{hyperref}

\usepackage[footnotesize]{subfigure}
\usepackage{amsthm}
\DeclareCaptionLabelSeparator{periodspace}{.\quad}
\usepackage{float}
\usepackage{cite}
\usepackage{epstopdf}

\usepackage[footnotesize]{subfigure}
\theoremstyle{remark}

\newcommand\ASTART{\bigskip\noindent\begin{minipage}[b]{0.5\linewidth}}
	
	\newcommand\AENDSKIP{\end{minipage}\bigskip}
\newcommand\AEND{\end{minipage}}
\ifCLASSOPTIONcaptionsoff
\usepackage[nomarkers]{endfloat}
\let\MYoriglatexcaption\caption
\renewcommand{\caption}[2][\relax]{\MYoriglatexcaption[#2]{#2}}
\fi
\theoremstyle{plain}

\newtheorem{lem}{\textbf{Lemma}}

\theoremstyle{definition}

\theoremstyle{remark}

\newcommand*{\rom}[1]{\expandafter\@slowromancap\romannumeral #1@}
\def\change{black}
\def\chang{black}
\def\changg{black}
\newcommand{\RN}[1]{%
\textup{\uppercase\expandafter{\romannumeral#1}}%
}

\usepackage{standalone}
\graphicspath{{./Figures/}}

\begin{document}
%
\title{Active User Detection and Channel Estimation for Spatial-based Random Access in Crowded Massive MIMO Systems via Blind Super-resolution}
\author{Abolghasem Afshar, Vahid Tabataba Vakili, Sajad Daei
	\thanks{The authors are with the school of Electrical Engineering, Iran University of Science and Technology}
}

\maketitle

\begin{abstract}
This work presents a novel framework for random access (RA) in crowded scenarios of massive multiple-input multiple-output (MIMO) systems. A huge portion of the system resources is dedicated as orthogonal pilots for accurate channel estimation which imposes a huge training overhead. This overhead can be highly mitigated by exploiting intrinsic angular domain sparsity of massive MIMO channels and the sporadic traffic of users, i.e., few number of users are active to send or receive data in each coherence interval. Besides, the continuous-valued angles of arrival (AoA) corresponding to each active user are alongside each other forming a specific cluster. To exploit these features in this work, we propose a blind clustering algorithm based on super-resolution techniques that not only detects the spatial features of the active users but also provides accurate channel estimation. Specifically, an off-grid atomic norm minimization is proposed to obtain the AoAs and then a clustering-based approach is employed to identify which AoAs correspond to which active users. After active user detection, an alternating-based optimization approach is performed to obtain the channels and transmitted data. Simulation results demonstrate the effectiveness of our approach in AoA detection as well as data recovery which indeed provides a high performance spatial-based RA in crowded massive MIMO systems.
\end{abstract}

%
\IEEEpeerreviewmaketitle

\section{Introduction}\label{section1}
 \IEEEPARstart{N}{owadays}, the number of wirelessly connected devices has been vastly increased by the development of applications such as Internet of Things (IoT), social networking and next generations of cellular communications including massive machine-type communications (mMTC), enhanced mobile broadband communications (eMBB) and ultra-reliable low-latency communications (URLLC).
 There are lots of advantages with massive MIMO systems such as increasing the system throughput and the energy efficiency \cite{massive2020,marzetta2010noncooperative}. As such, massive multiple-input multiple-output (MIMO) systems have received remarkable attention during the past few years. However, all of these advantages are dependent on accurate channel state information (CSI) in coherent transmission which is a highly challenging task. {\color{\chang}Due to the reciprocity of channel estimation (CE), Time Division Duplexing (TDD) mode is often preferred to Frequecny Division Duplexing (FDD) in MIMO systems. However, it requires a portion of resources as pilots or training signal in each Coherence Interval (CI) to estimate the channels corresponding to active users. In this regard, Random Access (RA) to pilots (RAP) is a promising solution for this pilot allocation which divides into two categories: grant-based and grant-free. In grant-based schemes (see \cite{afshar2021spatial,bjornson2017random}), multiple active user equipment (UE)s first transmit dedicated preambles selected from a pool of pilot sequences to access the Base Station (BS). Contention resolution schemes are then required if multiple users select the same pilots. Since large number of collisions occurs, the BS cannot resolve all of the contentions and thus many users are not able to access the BS.
\cite{bjornson2017random} proposes a strongest user collision resolution protocol to resolve this issue in crowded scenarios (which neglects the users in the edge of the cells) by using orthogonal pilots. \cite{afshar2021spatial} employs a deterministic Compressed Sensing (CS)-based RA scheme. As the signaling overhead and access latency is directly proportional to the number of users, conventional grant-based RA fails to support massive connectivity. In contrast, in grant-free RAP protocol, each active UE transmits pilots with embedded data without waiting for permission by BS \cite{massive2020,liu2018massive}. By using orthogonal pilots in this case, the number of active users that can access the BS is confined to the number of orthogonal pilots which is severely limited due to the short channel coherence time. Moreover, using non-orthogonal pilots in grant-free case complicates the task of active user detection (AUD) due to the intra- and inter-cell interference caused by correlated pilots \cite{ke2020compressive,chen2021sparse}. A large number of research works consider coordinated grant-free schemes in which the BS knows the pilots in advance (e.g. \cite{ke2020compressive,covarianced-based2020,djelouat2021joint,chen2019covariance}). As an example of this category, \cite{ke2020compressive} proposes a statistical Approximate Message Passing (AMP)-based algorithm for joint AUD and CE. Their method exploits the sporadic traffic of active users and the sparse nature of the channel but needs the full knowledge of the channels and noise distributions and assumes the angle of arrivals (AoAs) to be on a predefined domain of grids. Both of these assumptions are not generally satisfied in practice. Another line of research works lying in the coordinated grant-free category devotes to covariance-based activity identification which formulates the problem as an maximum likelihood estimation (see e.g. \cite{covarianced-based2020,chen2019covariance,chen2021phase,haghighatshoar2018improved}). It also has been shown that covariance-based algorithms outperforms AMP with the same length of the pilots \cite{covarianced-based2020}. Such kinds of schemes are based on the accuracy of the sample covariance of the measurement matrix. In order to have an accurate covariance, the number of required measurements has to be far more than the degrees of freedom (the true number of unknowns) and imposes a huge waste in bandwidth resources and cost. Overall, there are specific disadvantages with the mentioned prior works: full knowledge of channel and noise distribution, e.g. \cite{ke2020compressive,2018sparse}; the number of active users that can access the BS is severely limited, e.g. \cite{bjornson2017massive,ke2020compressive,covarianced-based2020}; A very large number of antennas and measurements are required, e.g. \cite{covarianced-based2020,haghighatshoar2018improved,chen2021sparse}. However, there is a common issue that the mentioned prior works are coordinated. This means that the BS has to be first know some pilots in advance for AUD which increases access latency, subsequently preventing to achieve a high spectral efficiency and seems to be not practical since the BS does know which user lies in which cell. It should be mentioned that \cite{zhang2017blind} provides a statistical uncoordinated data recovery and channel estimation (not necessarily designed for RA), however, besides its high complexity and the issue regarding the discrete nature of AoAs, it needs full knowledge of prior distributions of data and channel in advance which seems to be impractical.
To solve the mentioned challenges, we propose an uncoordinated grant-free RAP deterministic scheme in crowded scenarios of massive MIMO systems which leverages the intrinsic features of massive MIMO channels (angular domain sparsity of massive MIMO channels) as well as the sporadic traffic of users. The task of extracting such features from a few number of measurements builds upon the well-known framework of CS \cite{candes2006robust}}. Precisely, CS suggests a framework for recovering discrete-index parameters i.e. the unknown parameters are confined to be on a predefined domain of grids. There is also a more recent framework called continuous CS (or super-resolution) \cite{candes2014towards,tang2013compressed,sayyari2020blind,bayat2020separating} which assumes that the unknown parameters can lie anywhere and are not confined to be on the predefined grids. The aforementioned features are available in massive MIMO systems with massive number of users. For example, the physical channel of massive MIMO systems, employed in high frequencies (millimeter wave), have continuous sparse structure i.e. signal is received out of few off-grid (continuous) angles in BS antenna arrays and they can have any arbitrary values \cite{bajwa2010compressed}. This is due to the fact that signals with higher frequencies are more likely to be blocked by obstructions and few multi-path components (MPCs) contribute to the channel. Another feature is sporadic traffic of massive users which means only few users want to send their data at the same time. It has also been shown that the channels between users and BS exhibit a clustered continuous sparsity pattern \cite{ke2020compressive}. This implies that the AoAs corresponding to each user are alongside each other, few number of clusters are active and the AoAs have continuous-index values. {\color{\chang}For AUD and channel estimation, we design a deterministic optimization framework to encourage the mentioned features which does not need any distributions of channels and users' data. Our framework has three stages: first, we find the AoAs by solving an optimization problem, then a clustering-based algorithm is proposed to detach the angles corresponding to each active user and finally an alternative optimization algorithm is developed to estimate complex amplitudes of the channels and data/primary pilots transmitted by active users in a blind way. By our approach, the limitation in the number of adopted simultaneous users in crowded scenarios of massive MIMO systems would be resolved without any need for coordination between BS and users in advance.}
The organization of the paper is as follows: In Section \ref{sec.model}, the system model of massive MIMO is presented. Section \ref{sec.proposed} is about our proposed blind super resolution method and provides an algorithm for blind detection and CE. Lastly, Section \ref{sec.simulations} provides some numerical experiments to verify our proposed method. Lastly, the paper is concluded in Section \ref{sec.conclusion}.

  \textit{Notations}: 
We use boldface lower-and upper-case letters for vectors and matrices, respectively. The $i$-th element of a vector e.g. $\bm{x}$ and the $(i,j)$ element of a matrix e.g. $\bm{X}$ are respectively shown by $x_i$ and $X_{(i,j)}$. For vector $\bm{x}\in\mathbb{C}^n$ and matrix $\bm{X}\in\mathbb{C}^{n_1\times n_2}$, the $\ell_2$ norm and Frobenius norm are defined respectively as $\|\bm{x}\|_2:=({\sum_{i=1}^n|x(i)|^2})^{\tfrac{1}{2}}$ $\|\bm{X}\|_{F}:=\sqrt{\sum_{i=1}^{n_1}\sum_{j=1}^{n_2}|X(i,j)|^2}$. $\bm{X}\succeq \bm{0}$ means that $\bm{X}$ is a positive semidefinite matrix. For two arbitrary matrices $\bm{A}, \bm{B}$, $\langle \bm{A}, \bm{B}\rangle_R$ represents the trace of $\bm{B}^H\bm{A}$. $\mathcal{P}_{\Omega}(\cdot)$ is a operator transforming an arbitrary matrix to a reduced matrix with rows indexed by $\Omega$.
 \section{System Model}\label{sec.model}
We consider a typical uplink access scenario for MIMO-OFDM systems where there are one BS equipped with an $N$-element uniform linear array (ULA) and $K$ single-antenna users along with OFDM modulation to combat inter-symbol interference \cite[Section II]{ke2020compressive}. The sub-channel corresponding to each OFDM sub-carrier between the $k$-th user and the BS is modeled as (see \cite[Equ. 2]{ke2020compressive} or \cite[Equ. 7]{zhang2017blind}):
  \begin{align}
 \bm{h}_k=\sum_{l=1}^{L_k}\alpha_l^k\bm{a}(\theta_l^k)=\bm{A}_k\bm{\alpha}^k\in\mathbb{C}^{N\times 1},
 \end{align}  
 in which $L_k$ is the number of physical paths between $k$-th user and BS, $\theta_l^k$ is the Angle of Arrival (AoA) of the $l$-th path, $\alpha_{l}^k$ is the complex gain of the $l$-th path,
 \begin{align}
 \bm{a}_r(\theta)=\tfrac{1}{\sqrt{N}}[1, {\rm e}^{-j2\pi \Delta_r\cos(\theta)},...,{\rm e}^{-j2\pi \Delta_r (N-1)\cos(\theta)} ]^T
 \end{align}
 is the receive steering vector, $\bm{\alpha}^k:=[\alpha_1^k,..., \alpha_{L_k}^k]^T$, and $\bm{A}_k:=[\bm{a}_r(\theta_1^k),..., \bm{a}_r(\theta_{L_k}^k)]\in\mathbb{C}^{N\times L_k}$. Due the sparse characteristics of massive MIMO channels, it holds that $L_k\ll N$. The sub-channel is considered to be block fading, i.e., it is constant during several CIs where each is denoted by $T$. In each sub-carrier, the received signal at the BS after $T$ time slots at the sensors indexed by $\Omega \subseteq \{1,..., N\}$ (with length $|\Omega|:=M<N$) becomes in the form of (\cite{zhou2007experimental}, \cite[Equ. 3]{ke2020compressive} or \cite[Equ. 1]{zhang2017blind}):
 \begin{align}\label{eq.observed}
 \bm{Y}_{\Omega}=\mathcal{P}_{\Omega}(\bm{Y})=\mathcal{P}_{\Omega}(\sum_{k=1}^K\underbrace{\bm{h}_{k}\bm{s}_k^H)}_{:=\bm{X}_k}+\bm{W}\in \mathbb{C}^{M\times T},
 \end{align}  
 where $\bm{s}_k\in\mathbb{C}^T$ is the transmitted signal from $k$-th UE, $\bm{W}\in\mathbb{C}^{M\times T}$ is the additive noise matrix, each element of which is distributed as $\mathcal{CN}(0,\sigma^2)$ and $\bm{X}_k:=\sum_{l=1}^{L_k}\alpha_l^k\bm{a}(\theta_l^k)\bm{s}_k^{H}$. Inspired by \cite{chandrasekaran2012convex}, and by defining $c_l^k:=\alpha_l^k\|\bm{s}_k\|_2$ and $\bm{\phi}_k:=\tfrac{\bm{s}_k}{\|\bm{s}_k\|_2}$, $\bm{X}_k$ can be expressed as a sparse linear combinations of the matrix atoms in the atomic set ${ \mathcal{A}_k=\{\bm{a}_r(\theta)\bm{\phi}_k^H: \|{\bm{\phi}}_k\|_2=1, \theta\in (0,\pi)\},}$
 which are regarded as building blocks of $\bm{X}_k=\sum_{l=1}^{L_k}c_l^k\bm{a}(\theta_l^k)\bm{\phi}_k^{H}$. {\color{\chang}The aim is to extract the continuous parameters of $\bm{X}_k$ (i.e. the angles $\bm{\theta}_k$) by observing $\bm{Y}_{\Omega}$. Note that for inactive users, $\bm{\phi}_k$ and thus $\bm{X}_k$ are equal to zero.}
 \section{Proposed Blind Super-resolution method}\label{sec.proposed}
  In \eqref{eq.observed}, we have an under-determined set of equations with $N T$ observations and $K N T$ unknowns. While this problem has infinite number of solutions, it could be transformed to a tractable problem by assuming that $L_k \ll N$ which is reasonable in massive MIMO systems. This strategy is built upon well-known continuous CS approaches \cite{candes2014towards,tang2013compressed,valiulahi2019two} and provides a unique optimal set of solutions for matrices $\bm{X}_k$s in \eqref{eq.observed} leading to the least number of atoms under the affine constraints of \eqref{eq.observed}. Thus, we form the following optimization problem to reflect the structure of $\bm{X}_k$s:
 \begin{align}\label{prob.atomic_l0}
 &\min_{\substack{\bm{Z}_k\in\mathbb{C}^{N\times T}\\ k=1,..., K}} \sum_{k=1}^K \|{\bm{Z}_k}\|_{\mathcal{A}_k,0} ~s.t. \|\bm{Y}_{\Omega}-\sum_{k=1}^K\mathcal{P}_{\Omega}(\bm{Z}_k)\|_F\le \eta
 \end{align} 
 where $\|\bm{Z}_k\|_{\mathcal{A},0}:=\inf\big\{L_k: \bm{Z}_k=\sum_{l=1}^{L_k}c_{l}^k\bm{a}({\theta}_l^k)\bm{\phi}_k^H, c_l^k>0 , \bm{a}(\theta_l^k)\bm{\phi}_k^H\in\mathcal{A}_k\big\}
 $ is the atomic $\ell_0$ function which computes the least number of atoms to describe $\bm{Z}_k$. As \eqref{prob.atomic_l0} is an NP-hard problem in general, we relax  {\color{\chang} \eqref{prob.atomic_l0} into its closest convex optimization problem which is stated as}: 
  \begin{align}\label{prob.atomic_l1}
 &\min_{\substack{\bm{Z}_k\in\mathbb{C}^{N\times T} \\k=1,..., K}} \sum_{k=1}^K \|{\bm{Z}_k}\|_{\mathcal{A}_k} ~s.t.~\|\bm{Y}_{\Omega}-\sum_{k=1}^K\mathcal{P}_{\Omega}(\bm{Z}_k)\|_F\le \eta,
 \end{align}
 where {\color{\chang} the atomic norm $\|\cdot\|_{\mathcal{A}_k}$ is the best convex surrogate for the number of atoms composing $\bm{Z}_k$ (i.e. $\|\cdot\|_{\mathcal{A}_k,0}$) and is defined as the minimum of the $\ell_1$ norm of the coefficients forming $\bm{Z}_k$:}
 \begin{align}
 &\|\bm{Z}_k\|_{\mathcal{A}}:=\inf\{t>0: \bm{Z}_k\in t{\rm conv}(\mathcal{A}_k)\}=\nonumber\\
 &\inf\{\sum_{l=1}^{L_k}c_l^k: \bm{Z}_k=\sum_{l=1}^{L_k}c_{l}^k\bm{a}({\theta}_l^k)\bm{\phi}_k^H, c_l^k>0 , \bm{a}(\theta_l^k)\bm{\phi}_k^H\in\mathcal{A}_k\}
 \end{align}
 where ${\rm conv(\mathcal{A})}$ is the convex hull of $\mathcal{A}$. 
To identify the AoAs (which we used in our simulations) is by leveraging the solution of the dual problem of \eqref{prob.atomic_l1} which is provided below:
\begin{align}\label{prob.dual}
&\min_{\bm{V}\in\mathbb{C}^{N\times T},\bm{Z}\in\mathbb{C}^{T\times T}}2{\rm Re}\langle \bm{V}_{\Omega}, \bm{Y}_{\Omega}\rangle_F+2\eta\|\bm{V}_{\Omega}\|_F~~s.t. \nonumber\\
&
\begin{bmatrix}
\bm{I}&\bm{V}^H\\
\bm{V}&\bm{Z}_i
\end{bmatrix}\succeq \bm{0},\mathcal{T}^{*}(\bm{Z}_i)=\mathcal{T}^{*}(\bm{I}), i=1,..., K, ~\mathcal{P}_{\Omega}(\bm{V})=\bm{0}	
\end{align}
where $(\mathcal{T}^{*}(\bm{Z}))_k=\sum_{i=\max(1,k+1)}^{\min(k+T,T)}\bm{Z}_{i,i-k}, k=-(T-1),..., (T-1)$
is the adjoint operator of $\mathcal{T}$. Then, we use the following lemma (adapted from \cite[Lemma 1]{bayat2020separating} and \cite[Theorem 1]{sayyari2020blind}) which guarantees the uniqueness of the solution in the noiseless case:  
\begin{lem}\label{lem.uniuqeness}
	Denote the set of AoAs from $i$-th users by $\mathcal{S}_a^i=\{\theta_l^i\}_{l=1}^{L_i}$.
	The solutions of $\bm{Z}_k$ obtained from \eqref{prob.atomic_l1} in the noiseless case ($\eta=0$) are unique if there exist dual matrices $\bm{V}\in\mathbb{C}^{N\times T}$ such that the
	 vector-valued dual polynomials $\bm{q}_i(\theta)=\bm{V}^H\bm{a}_r(\theta)$
	satisfy the conditions
	\begin{align*}
	&\bm{q}_i(\theta)=\bm{\phi}_i, \forall \theta \in \mathcal{S}^i_a ,
		~\|\bm{q}_i(\theta)\|_2<1 \forall \theta \in [0,\pi)\setminus\mathcal{S}^i_a, i=1,..., K.
	\end{align*}
\end{lem}
This lemma shows that the AoAs can be easily estimated by identifying locations where $\|\bm{q}_i(\theta)\|_2$ achieves $1$. As stated in \cite{candes2013super}, this provides a good insight about the procedure of finding AoAs in the noisy case. Specifically, we find the AoAs by identifying the ones that $\|\bm{q}_i\|_2=1, i=1,..., K$.

After obtaining the estimated AOAs, a clustering-based algorithm \cite{kmeans} is employed to detect the AOAs of {\color{\chang} each cluster (active user)} denoted by $\bm{\theta}^{k}=[\theta_1^k,..., \theta^k_{\widehat{L}_k}]^T, k\in \widehat{\mathcal{S}}_u$. Here, {\color{\chang}$\widehat{\mathcal{S}}_u$ is the estimated set of indices corresponding to clusters (estimated active users) with known length $|\widehat{\mathcal{S}}_u|=K_a$}.
By knowing the AOAs corresponding to {\color{\chang}each cluster (estimated active user)}, \eqref{eq.observed} turns into the following equation:
 \begin{align}\label{eq.simpled_obs}
 \bm{Y}_{\Omega}=\sum_{k\in\widehat{\mathcal{S}}_u}\bm{A}^k_{\Omega}\bm{c}^k\bm{\phi}_k^H+\bm{W}_{M\times T},
 \end{align}
 where $\bm{A}^k_{\Omega}=\mathcal{P}_{\Omega}([\bm{a}_r(\theta^{k}_1),..., \bm{a}({\theta}^k_{\widehat{L}_k})])\in\mathbb{C}^{M\times K_a}$ is the steering matrix of $k$-th active user and $\bm{c}^k=[c_1^k,..., c^k_{\widehat{L}_k}]^T$. The task of recovering the unknown matrices $\bm{c}^k$ and $\bm{\phi}_k$ from $\bm{Y}_{\Omega}$ is a bi-linear inverse problem. For this task, we propose an alternating optimization to jointly estimate complex channel coefficients and transmitted data corresponding to active users. First, we begin with a random $\widehat{\bm{\phi}}_k$ distributed on the unit sphere. By replacing $\widehat{\bm{\phi}}_k$ in \eqref{eq.simpled_obs}, we deal with the following least square problem: 
 \begin{align}\label{eq.c_estimate}
 [\widehat{\bm{c}}^1,..., \widehat{\bm{c}}^{K_a}]=\mathop{\arg\min}_{\bm{c}^k,k=1,..., K_a}\|\bm{Y}_{\Omega}-\sum_{k\in\mathcal{S}_u}\bm{A}^k_{\Omega}\bm{c}^k\widehat{\bm{\phi}}_k^H\|_F
 \end{align}
 which can be easily solved by numerical optimization. 
By integrating the latter expression into \eqref{eq.simpled_obs}, we must solve the following least square optimization: ${ \widehat{\bm{\Phi}}=\mathop{\arg\min}_{\bm{\Phi}_{K_a\times T}}\|\bm{Y}_{\Omega}-\bm{B}\bm{\Phi}\|_F,}$
 where $\bm{B}:=[\bm{A}_{\Omega}^1\widehat{\bm{c}}^1,...,\bm{A}_{\Omega}^{K_a}\widehat{\bm{c}}^{K_a} ]\in\mathbb{C}^{M\times K_a}$
 and $\bm{\Phi}:=[\bm{\phi}_1,..., \bm{\phi}_{K_a}]^T$. The latter optimization has also the closed-form solution
 \begin{align}\label{eq.phi_estimate}
[\widehat{\bm{\phi}}_{1},..., \widehat{\bm{\phi}}_{K_a}]^T=\widehat{\bm{\Phi}}=\bm{B}^{\dagger}\bm{{Y}}_{\Omega}.
 \end{align}
 Finally, the steps \eqref{eq.c_estimate} and \eqref{eq.phi_estimate} are alternatively performed to yield the final solution. The pseudo code of the proposed method which is indeed a summary of the aforementioned steps is provided in Algorithm \ref{algorithm.admm}.
{\centering
\resizebox{.5\textwidth}{!}{
\begin{minipage}{.8\textwidth}
\begin{algorithm}[H]
	\caption{}
	\begin{algorithmic}[1]\label{algorithm.admm}
		\REQUIRE $\bm{Y}\in\mathbb{C}^{M\times T}$,$\eta$, $K_a$, maxiter
	\STATE Select a uniformly distributed random vector for transmitted data as $\widehat{\bm{\phi}}_k=5 rand(T,1) \forall~ k=1 ~\text{to}~ K_a$
	\STATE $\bm{A}_{total}=\emptyset$
			\begin{itemize}
		\item Solve the dual problem \eqref{prob.dual} to obtain $\bm{V}$ as follows:
	 \item Obtain the dual polynomial $\bm{q}_i(\theta)=\bm{V}^H\bm{a}_r(\theta), i=1,..., K$.
		
	 \item Localize the estimated angle $\widehat{\theta}\in [0,1]$ by the following two methods:
	
	\item	Discretize $\widehat{\theta}$ on a fine grid up to a desired accuracy and find $\widehat{\theta}$ and by identifying locations where $\|\bm{q}_i(\theta)\|_2, i=1,..., K$ achieves to $1$ according to Lemma \ref{lem.uniuqeness}. The total number of angles reaching $1$ specifies an estimate for the total number of MPCs i.e. $\sum_{k=1}^{K_a}L_k$ 
	\end{itemize}
\item
\begin{itemize}
	\end{itemize}

	\STATE Apply the k-means methods to cluster the angles of channel UEs.
	
	\STATE $[{\rm label}]={\rm k-means}(\widehat{\theta},K_a)$
	\FOR{$k=1$ to $K_a$}
	\STATE Identify the corresponding indices with the $k$-th label.
	\STATE Estimate the length of $k$-th cluster i.e. $L_k$.
\STATE Obtain the angles corresponding to the $k$-th cluster (UE) i.e. $\widehat{\theta}_1^k,..., \widehat{\theta}_{L_k}^k$
\STATE Estimate the steering matrix as
\STATE $\bm{A}_{total}\leftarrow [\bm{A}_{total}, \bm{A}_k]$
	\ENDFOR
	\FOR {$i=1$ to maxiter}
	\STATE Recover $[\widehat{\bm{c}}^1,..., \widehat{\bm{c}}^{K_a}]$ according to \eqref{eq.c_estimate}.
	\STATE Recover $[\widehat{\bm{\phi}}_1,..., \widehat{\bm{\phi}}_{K_a}]$ according to \eqref{eq.phi_estimate}.
	\STATE $\widehat{\bm{\phi}}_k\leftarrow  \tfrac{\widehat{\bm{\phi}}_k}{\|\widehat{\bm{\phi}}_k\|_2}$.
	\ENDFOR
	\end{algorithmic} 
	Return: $ \widehat{\bm{\theta}}^k,\widehat{ \bm{\alpha}}^k, \widehat{ \bm{s}}_k, ~ \forall k \in\mathcal{S}_u$. 
\end{algorithm} 
\end{minipage}
}
}
\textit{Discussion}:
Our algorithm provides a blind spatial-based RA scheme which simultaneously estimates data as well as channels. The only assumption on the data to be unambiguously recovered is positivity i.e. $\bm{s}_i>0$ and normalized power $\|\bm{s}_i\|_2=1$ for all active users $i=1,..., K_a$. {\color{\chang}The proposed method has implications for data recovery in mMTC as well as RAP and AOA detection in crowded mobile broadband communications (cMBB).} For example, CI in cMBB divides into 2 parts: RAP and data transmission blocks. By utilizing this novel approach in the RAP block, AUD and AoA estimation are performed by the BS \cite[Section 5]{afshar2021spatial}. By our method, BS can identify many users at the same time with their AoAs and with the lowest level of spending system resources. In fact, the number of active users that can access the network depends on the complexity that BS can bear. There is no need for orthogonality of RA pilots needed for RAP process in \cite{bjornson2017random} and prior distribution of pilots and channels as is the case in AMP-based approaches \cite{zhang2017blind,ke2020compressive}. {\color{\change}After RAP, in the coherent transmission step, BS can easily allocate dedicated orthogonal pilots to non-overlapped UEs and estimates their corresponding data via \eqref{eq.phi_estimate} by knowing the exact AoAs of active users in the RAP stage.}
 \section{Simulations}\label{sec.simulations}
\begin{figure}[t]
	\hspace{-.6cm}
	\includegraphics[scale=0.28]{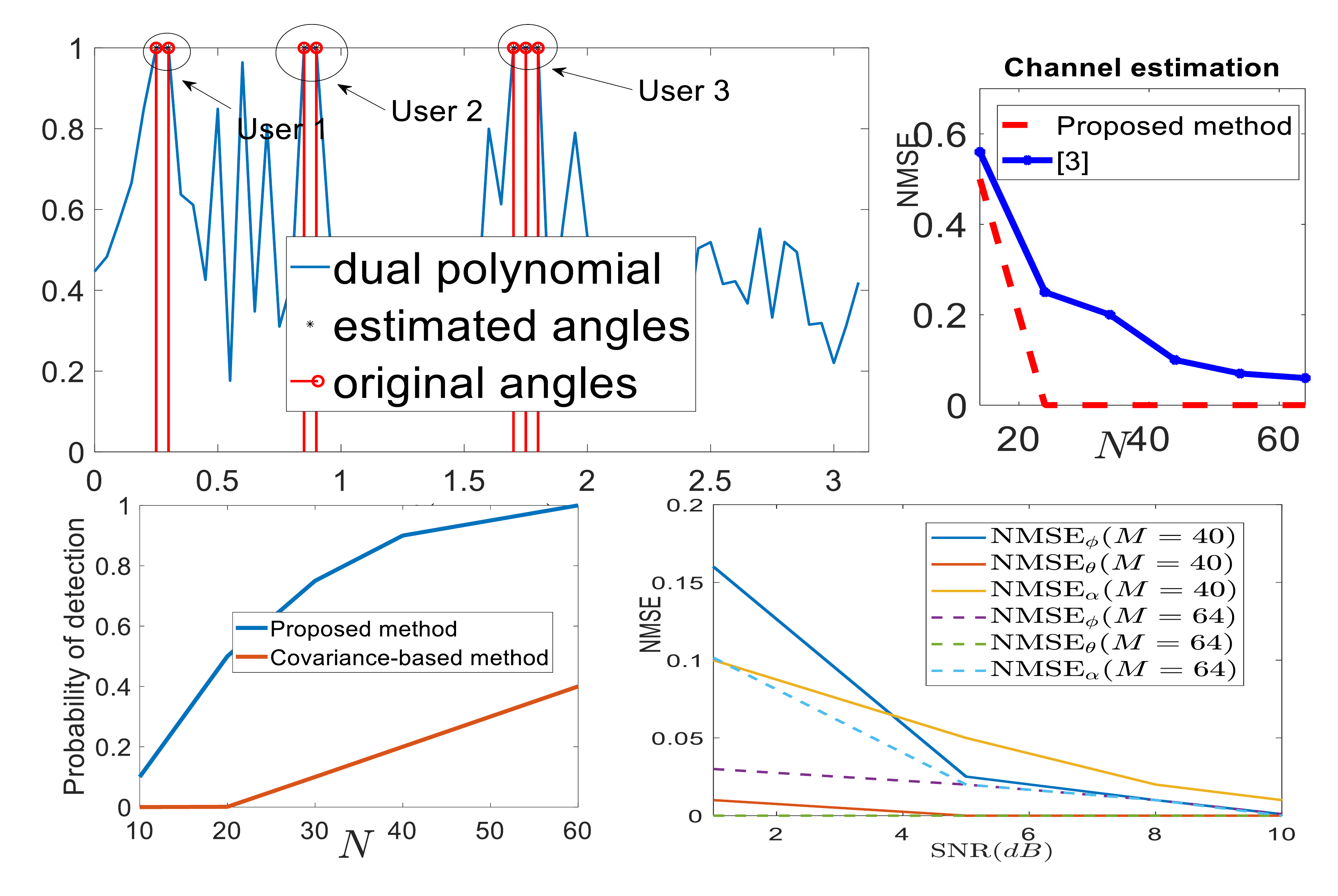}
	\caption{Top-left image: This image depicts $\ell_2$ norm of the dual polynomial vector. One can find the angles of active users by identifying angles with maximum amplitude. The angles of active user channels are clustered using kmeans method. The used parameters are $N=64, M=30, K_a=3, K=10, T=2, SNR=10~dB.$ Top-right image: This image compares the performance of our algorithm in CE with \cite{afshar2021spatial} for different number of BS antennas with settings $L_{\max}=4, T=10, K_a=3, K=10, {\rm SNR}=3dB $. Bottom left image: {\color{\changg}The performance of AUE is compared with \cite{haghighatshoar2018improved} versus the number of observed arrays with parameters $N=60, M=60, K_a=5, K=50, L_{\max}=3, {\rm SNR}=5~dB$}. Bottom right image: This image shows the performance of our algorithm in estimating angles, pilots and complex amplitudes with parameters $N=64, L_{\max}=3, K_a=12, K=40, T=10$.}\label{fig.nmse}
\end{figure}

 In this section, we perform some numerical experiments to evaluate the performance of our proposed algorithm in blind channel and data reconstruction. We use SDPT3 package of CVX \cite{cvx} in MATLAB for solving problem \eqref{prob.dual}. The number of BS antennas is set to $N=64$. We assume the one-ring model for the channel \cite{nam2014joint}. The AoAs are randomly chosen from $[0,\pi]$. The path amplitudes are distributed as $\mathcal{CN}({0},1)$. The separation between any receive antennas at BS are set to $\Delta_r=0.5$. The observed sensors at BS ($\Omega$) are randomly chosen out of $\{1,..., N\}$. Also, the maximum number of iterations in Algorithm \ref{algorithm.admm} denoted by maxiter is fixed to $5$. The upper bound of noise variance is chosen as $\eta=\|\bm{W}\|_F$. The signal to noise ratio is defined by ${\rm SNR}=10\log_{10}(\frac{\|\mathcal{P}_{\Omega}(\bm{X})\|_F^2}{MT\sigma^2})$. First, in the top-left image of Figure \ref{fig.nmse}, we show the successful procedure of AUE and clustering with parameters $M=30$, $K=10$, $K_a=3$, $T=2$, ${\rm SNR}=10~dB$. The maximum number of MPCs is fixed to $L_{\max}=3$. This image shows the $\ell_2$ norm of the vector-valued dual polynomial at different angles in terms of radian. The estimated angles are found by identifying locations that $\|\bm{q}(\theta)_i\|_2=1, i=1, ..., K$. The number of peaks provides an estimate for $\sum_{k\in\mathcal{S}_u}L_k$. After finding the angles, we apply k-means method to cluster the angles corresponding to $K_a$ active users. The number of elements inside each cluster provides an estimate for $\widehat{ L}_k$. Then, steps 12 to 16 of Algorithm \ref{algorithm.admm} are employed to obtain the pilots and channels. The performance of our algorithm in recovering users' data, channel amplitudes and AoAs is evaluated using normalized mean square error (NMSE) respectively defined by ${\rm NMSE}_{\bm{\phi}}:=\mathds{E}\sqrt{\frac{\sum_{k=1}^{K_a}\|\bm{\phi}_k-\widehat{\bm{\phi}}_k\|_2^2}{\sum_{k=1}^{K_a}\|\bm{\phi}_k\|_2^2}}$, ${\rm NMSE}_{\bm{\alpha}}:=\mathds{E}\sqrt{\frac{\sum_{k=1}^{K_a}\|\bm{\alpha}_k-\widehat{\bm{\alpha}}_k\|_2^2}{\sum_{k=1}^{K_a}\|\bm{\alpha}_k\|_2^2}}$ and ${\rm NMSE}_{\bm{\theta}}:=\mathds{E}\sqrt{\frac{\sum_{k=1}^{K_a}\|\bm{\theta}^k-\widehat{\bm{\theta}}^k\|_2^2}{\sum_{k=1}^{K_a}\|\bm{\theta}^k\|_2^2}}$. The Monte-Carlo iterations to approximate the expectation is set to $50$. The evaluation for the first experiment are as follows: ${\rm NMSE}_{\bm{\Phi}}=10^{-6}, {\rm NMSE}_{\bm{\alpha}}=10^{-5}, {\rm NMSE}_{\theta}=10^{-8}$. In the second experiment, we evaluate the performance of Algorithm \ref{algorithm.admm} in different noise values in a more practical scenario with parameters $N=64, L_{\max}=3, K_a=12, K=40, T=10$. As it turns out from the bottom image of bottom-right image of Figure \ref{fig.nmse}, NMSEs tends to zero at high SNRs which in turn implies that our proposed method performs well in estimating users' data, complex channel amplitudes and AoAs of active users. In the experiment shown in the top-right image of Figure \ref{fig.nmse}, we compare our method with \cite{afshar2021spatial} for different number of antennas. For both methods, we obtain NMSE of the channel matrix defined by ${\rm NMSE}_{\bm{h}}:=\mathds{E}\sqrt{\frac{\sum_{k=1}^{K_a}\|\bm{h}_k-\widehat{\bm{h}}_k\|_2^2}{\sum_{k=1}^{K_a}\|\bm{h}_k\|_2^2}}$. As it can be observed, our blind method performs better in CE than \cite{afshar2021spatial} which assumes the users' data known. In the last experiment, {\color{\changg}the performance of AUE in our algorithm is compared with \cite[Algorithm 1]{haghighatshoar2018improved}} by a detection rate criterion defined as $DR=\frac{|\mathcal{S}_u-\widehat{\mathcal{S}_u}|}{K_a}$ where the numerator returns the number of differences between the true active users and the estimates. As shown in the bottom-left image of Figure \ref{fig.nmse}, the probability of detection enhances by increasing the {\color{\changg} number $N$ of BS antennas}.
 \section{Conclusion}\label{sec.conclusion}
 In this work, we designed a novel blind spatial-based random access solution which is applicable to crowded massive MIMO systems. Specifically, we showed that the recovery of both pilots and AoAs are possible via observing a few noisy measurements in blind manner. For this task, we used a clustering method to demix the AoAs corresponding to active user and an alternating-based approach is designed to recover the pilots and the complex amplitudes of the channels.

\ifCLASSOPTIONcaptionsoff

\fi
\bibliographystyle{ieeetr}
\bibliography{references1}
 \end{document}